\documentclass[a4paper,11pt]{article}
\usepackage{pos}

\usepackage{subcaption}

\title{Quantum computing for lattice supersymmetry}

\author*{Christopher Culver}
\author{David Schaich}

\affiliation{Department of Mathematical Sciences, University of Liverpool,\\
  Liverpool L69 7ZL, United Kingdom}

\emailAdd{C.Culver@liverpool.ac.uk}
\emailAdd{David.Schaich@liverpool.ac.uk}

\abstract{Quantum computing promises the possibility of studying the real-time dynamics of non-perturbative quantum field theories while avoiding the sign problem that obstructs conventional lattice approaches. Current and near-future quantum devices are severely limited by noise, making investigations of simple low-dimensional lattice systems ideal testbeds for algorithm development. Considering simple supersymmetric systems, such as supersymmetric quantum mechanics with different superpotentials, allows for the analysis of phenomena like dynamical supersymmetry breaking. We present ongoing work applying quantum computing techniques to study such theories, targeting real-time dynamics and supersymmetry breaking effects.}

\FullConference{%
  The 38th International Symposium on Lattice Field Theory, LATTICE2021 \\
  26--30 July 2021 \\
  Zoom/Gather @ Massachusetts Institute of Technology
}

\newcommand{\de}{\ensuremath{\delta} }

\newcommand{\bhat}{\ensuremath{\hat b} }
\newcommand{\bhatdag}{\ensuremath{\hat b^{\dag}} }

\newcommand{\phat}{\ensuremath{\hat p} }
\newcommand{\qhat}{\ensuremath{\hat q} }

\newcommand{\Qbar}{\ensuremath{\overline Q} }
\newcommand{\cW}{\ensuremath{\mathcal W} }

\newcommand{\ket}[1]{\ensuremath{\left| #1 \right\rangle} }
\newcommand{\zero}{\ensuremath{\ket{0}} }
\newcommand{\one}{\ensuremath{\ket{1}} }

\newcommand{\matElem}[2]{\ensuremath{\left| #1 \right\rangle \left\langle #2 \right|}} 

\newcommand{\Tr}[2]{\ensuremath{\mbox{Tr}_{#1} \left[ #2 \right]} }

\newcommand{\eq}[1]{Eq.~\ref{#1}}

\newcommand{\mysection}[1]{\vspace{-6 pt}\section{#1}\vspace{-6 pt}}
\newcommand{\mysubsection}[1]{\vspace{-6 pt}\subsection{#1}\vspace{-6 pt}}

\newcommand{\ahat}{\ensuremath{\hat a}}
\newcommand{\ahatdag}{\ensuremath{\hat{a}^{\dagger}}}

\usepackage{mleftright,xparse}
\NewDocumentCommand\xDeclarePairedDelimiter{mmm}
{%
	\NewDocumentCommand#1{som}{%
		\IfNoValueTF{##2}
		{\IfBooleanTF{##1}{#2##3#3}{\mleft#2##3\mright#3}}
		{\mathopen{\csname##2\endcsname#2}##3\mathclose{\csname##2\endcsname#3}}%
	}%
}
\xDeclarePairedDelimiter{\av}{\langle}{\rangle}
\xDeclarePairedDelimiter{\abs}{\lvert}{\rvert}
\NewDocumentCommand\braket{somm}{%
	\IfNoValueTF{#2}{\mleft\langle #3\,|#4\mright\rangle}{NOTIMPLEMENTED}
}
\NewDocumentCommand\opbraket{sommm}{%
	\IfNoValueTF{#2}
	{\IfBooleanTF{#1}{\langle#3|#4|#5\rangle}{\mleft\langle #3 \left| #4 \right| #5 \mright\rangle}}
	{\mathopen{\csname#2\endcsname\langle}#3\mathopen{\csname#2\endcsname|} #4 \mathclose{\csname#2\endcsname|} #5\mathclose{\csname#2\endcsname\rangle}}
}

\begin{document}
\maketitle

\mysection{Introduction}
Supersymmetry is an extension of Poincar\'e symmetry to include spinor generators relating bosonic and fermionic fields.  Supersymmetric theories have rich features including appearance in holographic dualities~\cite{Itzhaki:1998dd} which are a correspondence between spacetime in quantum gravity and a quantum field theory on its boundary.  The existence of dualities enables the indirect study of phenomena in quantum gravity, such as the dynamics of evaporating black holes.  Instead of solving the quantum gravity problem it is sufficient to study the quantum field theory.  In this realm lattice regularization provides a non-perturbative approach via both importance-sampling algorithms in classical computing as well as the developing frontier of quantum simulation.
Here we study simple supersymmetric systems using quantum computing, which in principle offers direct access to real-time dynamics of quantum field theories.
Beyond this, quantum computing may enable the investigation of entanglement generation~\cite{Buividovich:2018scl}, dynamical supersymmetry breaking (which has a severe sign problem~\cite{Bergner:2016sbv, Schaich:2018mmv}), and real-time scattering of particles in supersymmetric extensions of the standard model.

Quantum computing is currently in the Noisy Intermediate Scale Quantum~(NISQ) era~\cite{preskill2018quantum} of quantum devices with low qubit counts and significant error rates for the operations on qubits.  The solution to an evaporating black hole or real-time scattering are beyond the capabilities of NISQ devices.  Modern quantum computers are best suited to validating algorithmic advances on simple test case systems.  Here we will study simple supersymmetric quantum mechanics (SQM) with two supercharges in $0+1$ dimensions.  This system was recently studied with classical-computing lattice methods by Refs.~\cite{Kadoh:2018ivg,Joseph:2020gdh, Dhindsa:2020ovr}, which explored dynamical supersymmetry breaking for different superpotentials.
Similar supersymmetric matrix models are also under consideration as targets for quantum computing~\cite{Gharibyan:2020bab, Buser:2020cvn, Rinaldi:2021jbg}.
In this work we will study dynamical supersymmetry breaking for three different superpotentials.  In Section~\ref{sec:sqm} we define SQM in more detail and discuss how to test whether supersymmetry is spontaneously broken.  Then in Section~\ref{sec:qc} we discuss the necessary quantum computing technology to analyze our supersymmetric system.  Our progress is presented in Section~\ref{sec:progress} including determinations of supersymmetry breaking and the feasibility of studying SQM on NISQ devices.

\mysection{Supersymmetric Quantum Mechanics}\label{sec:sqm}
In this section we review SQM with two supercharges in $0+1$ dimensions.  We start by defining the Hamiltonian
\begin{equation}
  \label{eq:SQM}
  2H_{\text{SQM}} = 2i\partial_t = \left\{Q, \Qbar\right\} = Q \Qbar + \Qbar Q,
\end{equation}
where $Q$ and \Qbar are two independent supercharges with $Q^2 = \Qbar^2 = 0$ and $[H, Q] = [H, \Qbar] = 0$.  These can be expressed in terms of bosonic and fermionic operators as
\begin{align}
  Q     & = \bhat\left[i\phat + W'(\qhat)\right] &
  \Qbar & = \bhatdag\left[-i\phat + W'(\qhat)\right],
\end{align}
where \qhat and \phat are the bosonic coordinate and conjugate momentum operators respectively.  The function $W(\qhat)$ is the superpotential, and the prime denotes derivatives with respect to $\qhat$.
The fermionic operators $\hat{b}^{\dagger}$ and $\hat{b}$ create and destroy a fermionic state respectively, with $\bhat^2 = (\bhatdag)^2 = 0$ and $\left\{\bhat, \bhatdag\right\} = 1$.

Using the above equations it is instructive to write down the Hamiltonian in terms of bosonic and fermionic operators, for which
\begin{align}
  2H_{\text{SQM}}=\phat^2 + [W'(\qhat)]^2 - W''(\qhat) \left[\bhatdag, \bhat\right].
\end{align}
The coupling between the fermion and boson depends entirely on the superpotential.
The Hilbert space of the Hamiltonian is the tensor product of the Hilbert space of a single boson with that of a single fermion.  The continuous bosonic Hilbert space can be written with the basis $\ket{\phi}$ where
\begin{align}
  \left\langle \phi' \mid \phi \right\rangle & = \delta(\phi'-\phi), &
  \qhat\ket{\phi} & =\phi\ket{\phi}, &
  e^{ip\Delta}\ket{\phi} & =\ket{\phi+\Delta}.
\end{align}
The fermionic Hilbert space is a two-dimensional vector space spanned by $\ket{0}$ and $\ket{1}$ with
\begin{align}
  \bhat \one  & = \zero & \bhatdag \one  & = 0 \cr
  \bhat \zero & = 0     & \bhatdag \zero & = \one.
\end{align}
The bosonic Hilbert space needs to be regulated to map the degrees of freedom to a finite number of qubits, which we will discuss in Section~\ref{sec:qc}.

In this work we consider superpotentials for which the bosonic part of the Hamiltonian can be written as a harmonic oscillator plus additional interactions, $[W']^2 \supset m^2 \qhat^2$, where $m$ is the mass of the particles.
It is thus useful to transform from the $\qhat,\phat$ basis to the $\ahatdag,\ahat$ creation and annihilation operator basis of a harmonic oscillator, through the usual definitions
\begin{align}
  \ahat & = \sqrt{\frac{m}{2}}\qhat + \frac{i}{\sqrt{2m}}\phat, &
  \ahatdag & = \sqrt{\frac{m}{2}}\qhat - \frac{i}{\sqrt{2m}}\phat.
\end{align}

The SQM Hamiltonian in \eq{eq:SQM} is manifestly supersymmetric, but supersymmetry may break spontaneously depending on the superpotential.
One way to study dynamical supersymmetry breaking is to compute the Witten index~\cite{Witten:1981nf}
\begin{equation}
  \cW = \Tr{}{(-1)^F e^{-i H t}} = \Tr{B}{e^{-i H t}} - \Tr{F}{e^{-i H t}}.
\end{equation}
Here the first trace over all states of the Hilbert space is split into traces over the bosonic and fermionic parts of the Hilbert space.  A vanishing Witten index, $\cW = 0$, is a necessary but insufficient condition for supersymmetry breaking, and corresponds to a vanishing partition function $Z = 0$ that obstructs classical importance-sampling computations~\cite{Bergner:2016sbv, Schaich:2018mmv}.
Turning to quantum computing, a more straightforward approach is to exploit the fact that supersymmetry is broken only if the ground-state energy is non-zero.  Computing the ground-state energy is a task well suited for the variational quantum eigensolver~(VQE) to be discussed in the next section.

\mysection{Quantum Computing}\label{sec:qc}
To enable computations on a quantum computer we must regulate the infinite bosonic degrees of freedom.  We impose a hard cutoff $\Lambda$ of the boson by limiting the number of accessible modes of the harmonic oscillator.
While this explicitly breaks supersymmetry, we expect that these effects will become negligible for sufficiently large $\Lambda$.
The raising and lowering operators become
\begin{align}
  \ahat & = \sum_{n=0}^{\Lambda-2}\sqrt{n+1}\matElem{n}{n+1}, &
  \ahatdag & = \sum_{n=0}^{\Lambda-2}\sqrt{n+1}\matElem{n+1}{n}.
\end{align}
Having a Hilbert space with a finite number of degrees of freedom, we can convert the harmonic oscillator degrees of freedom to qubit ones.  The first step is to write the state $j$ in binary as $j=\sum_{i=0}^{N-1}b_i2^i$ with $N=\lceil\log_2\Lambda\rceil$.
We associate one qubit with each binary digit, and write the state as a tensor product of $N$ qubit states,
\begin{equation}
    \ket{j}=\ket{b_0}\ket{b_1}\ldots\ket{b_{N-1}}.
\end{equation}
Using this, we can translate the matrix element of any bosonic operator into its representation in terms of the qubit degrees of freedom,
\begin{equation}
  \ket{n}\langle n' \mid = \otimes_{i=0}^{N-1}\matElem{b_i}{b_i'}.
\end{equation}
It is useful to write the tensor product on the right hand side in terms of Pauli matrices using the following four relations,
\begin{align}
  \matElem{0}{1} & =\frac{1}{2}\left(X+iY\right), &
  \matElem{1}{0} & =\frac{1}{2}\left(X-iY\right), \\
  \matElem{0}{0} & =\frac{1}{2}\left(1+Z\right), &
  \matElem{1}{1} & =\frac{1}{2}\left(1-Z\right).
\end{align}
Any bosonic operator can thus be expressed as a linear combination of products of Pauli matrices, called a Pauli string.

For the fermionic degrees of freedom we use the Jordan--Wigner transformation which maps the fermionic occupation to spin degrees of freedom.  The creation and annihilation operators are
\begin{align}
  \bhatdag & = \frac{1}{2}\left(X-iY\right), &
  \bhat & = \frac{1}{2}\left(X+iY\right).
\end{align}
We will always assign the fermionic degree of freedom to the $N+1$ qubit.  The mapping from bosonic and fermionic degrees of freedom to qubit ones results in the Hamiltonian as a single Pauli string.  This allows us to take advantage of IBM Qiskit~\cite{treinish_2021_5154034} implementations of quantum algorithms such as Trotter evolution and the VQE algorithm.
We conclude this section by briefly reviewing each of these algorithms in turn.

First, to study real-time dynamics we need a quantum circuit for the operation $e^{iHt}$ acting on the qubits.  This can be done with the Suzuki--Trotter decomposition which breaks the continuous time $t$ into $N$ `Trotter steps' of size $\de \equiv t/N$ via
\begin{equation}
    e^{-i H t} \ket{\psi} = \left(\exp\left[-i H \de\right]\right)^N \ket{\psi}.
\end{equation}
For a Hamiltonian that is a sum of $M$ terms, $H=\sum_{j=1}^M H_j$, a single step becomes
\begin{equation*}
  \ket{\psi(t + \de)} \equiv \exp\left[-i \sum_{j = 1}^M H_j \de\right] \ket{\psi(t)}.
\end{equation*}
In practice $M > 1$ and the Baker--Campbell--Hausdorff formula is used to convert the exponential of the sum $H$ into a product of exponentials involving individual $H_j$.
Since these various terms in the Hamiltonian do not commute in general, this implies a trade-off between the number of gate operations and the accuracy with which the state $\ket{\psi(t+\de)}$ can be obtained at the end of a single Trotter step.
There is a similar trade-off when we consider that the Suzuki--Trotter decomposition also introduces errors dependent on $\de$ and on the ordering of gate operations within the circuit.  A simple evolution step such as that considered in Ref.~\cite{Brower:2020huh} reduces gate counts at the cost of larger errors, while more complicated evolution steps can reduce errors at the cost of (typically significant) increases in the number of gate operations.  In Section~\ref{sec:progress} we discuss gate costs for a single Trotter step for SQM with various superpotentials.

The VQE is a hybrid quantum--classical algorithm that finds the lowest eigenvalue of a matrix~\cite{McClean_2016}.
When the matrix in question is a Hamiltonian, this provides an approximation to the ground state as the corresponding eigenvector.
The algorithm functions similarly to the variational method.  First a trial wave function $\psi$ for the ground state is created which is defined with tunable parameters $\theta_i$.  A quantum circuit is used to compute the energy of the parameterized wave function, whose parameters are optimized via classical methods to minimize this energy.  Eventually the algorithm will converge to a value $E_{\text{var}}$ that is an upper bound for the ground-state energy,
\begin{equation}
  E_0\leq E_{\text{var}}=\frac{\opbraket{\psi(\theta_i)}{H}{\psi(\theta_i)}}{\braket{\psi(\theta_i)}{\psi(\theta_i)}}.
\end{equation}

In this work we report on simulations of VQE analyses using classical computing rather than actual quantum hardware.
This leaves the performance of the VQE dependent on the choice of trial wave function and classical optimizer.  For the trial wave function we use the $R_y$ variational form~\cite{treinish_2021_5154034}, which only involves $y$ rotations and CX gates.  For the classical optimizer we use COBYLA which was shown to work well in Ref.~\cite{Rinaldi:2021jbg}.  In the next section we present results for different superpotentials and cutoffs.  Given that the VQE provides only an upper bound on the ground-state energy, we perform $100$ runs for each computation and quote the minimum result as the best approximation to the ground state.

\mysection{VQE analyses for various superpotentials}
\label{sec:progress}
Here we discuss the current status of our studies of SQM with three different superpotentials, for a range of cutoffs $\Lambda$.  For each superpotential we start by looking at the eigenvalues of the Hamiltonian via exact diagonalization.  Then we compare these exact values to the best result obtained from $100$ runs of the VQE.  Lastly we discuss the cost of attempting to study the real-time dynamics of such systems in terms of their entangling gate counts.

\begin{figure}
    \centering
    \includegraphics[width=0.45\linewidth]{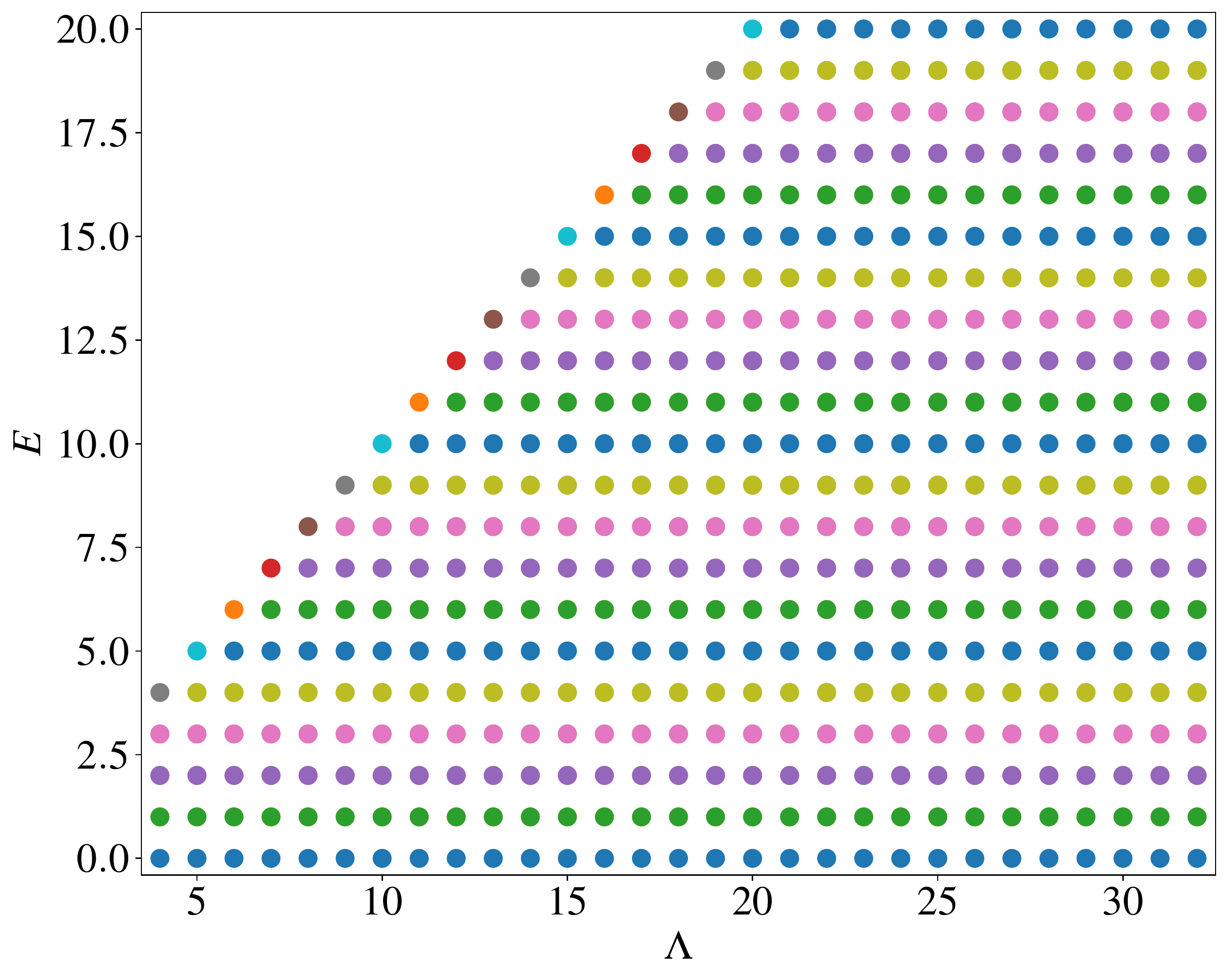}
    \hfill
    \includegraphics[width=0.45\linewidth]{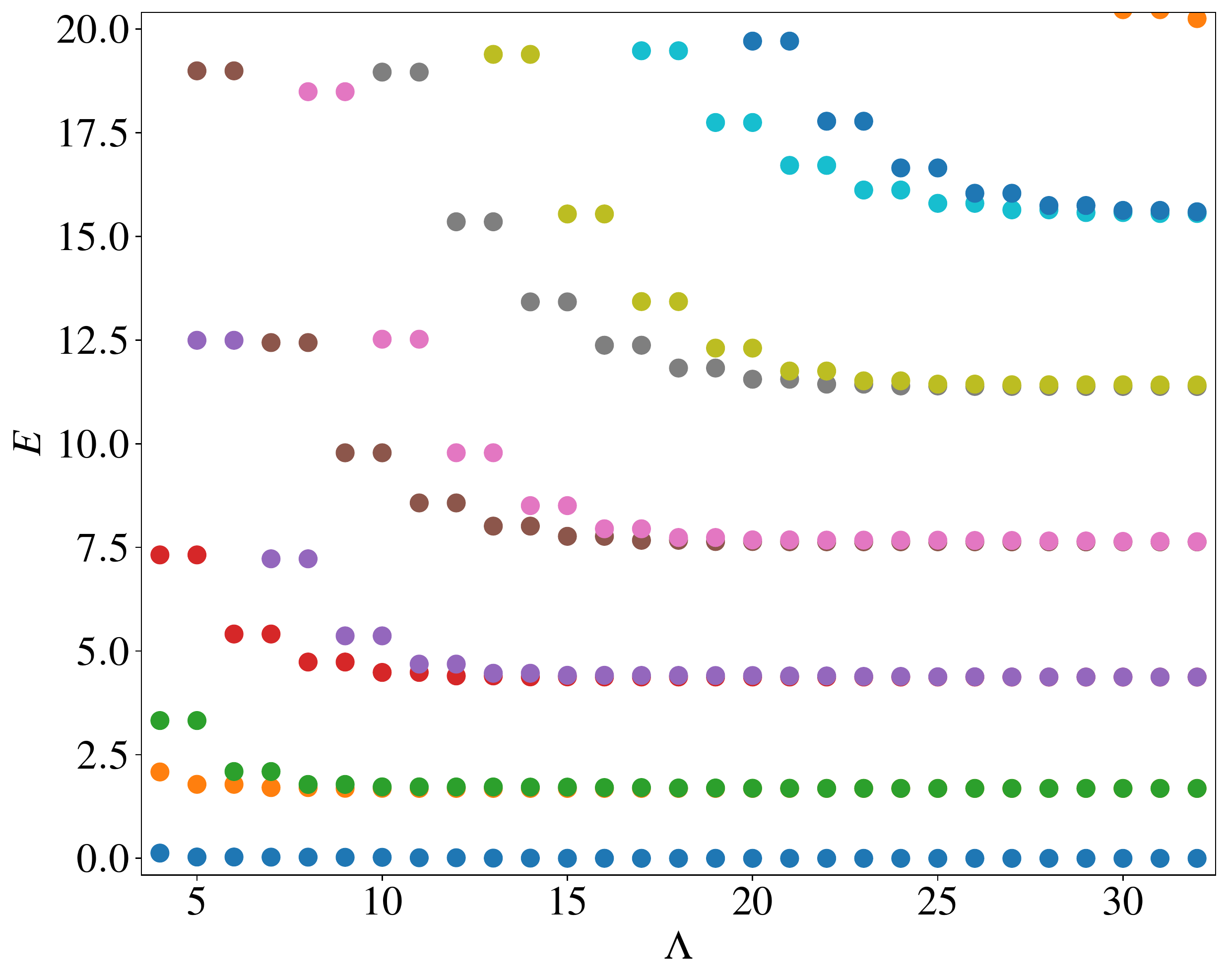}
    \caption{Spectra for different cutoffs of the bosonic degrees of freedom, for the harmonic oscillator (\textbf{left}) and anharmonic oscillator (\textbf{right}) superpotentials (with $m=g=1$).  For each $\Lambda$ there are $2\Lambda$ states which are differentiated by color.  For the harmonic oscillator, and the large-cutoff limit of the anharmonic oscillator, all levels except the ground state and highest-energy state are two-fold degenerate.}
    \label{fig:ho_spectrum}
\end{figure}

\begin{figure}
    \centering
    \includegraphics[width=0.45\linewidth]{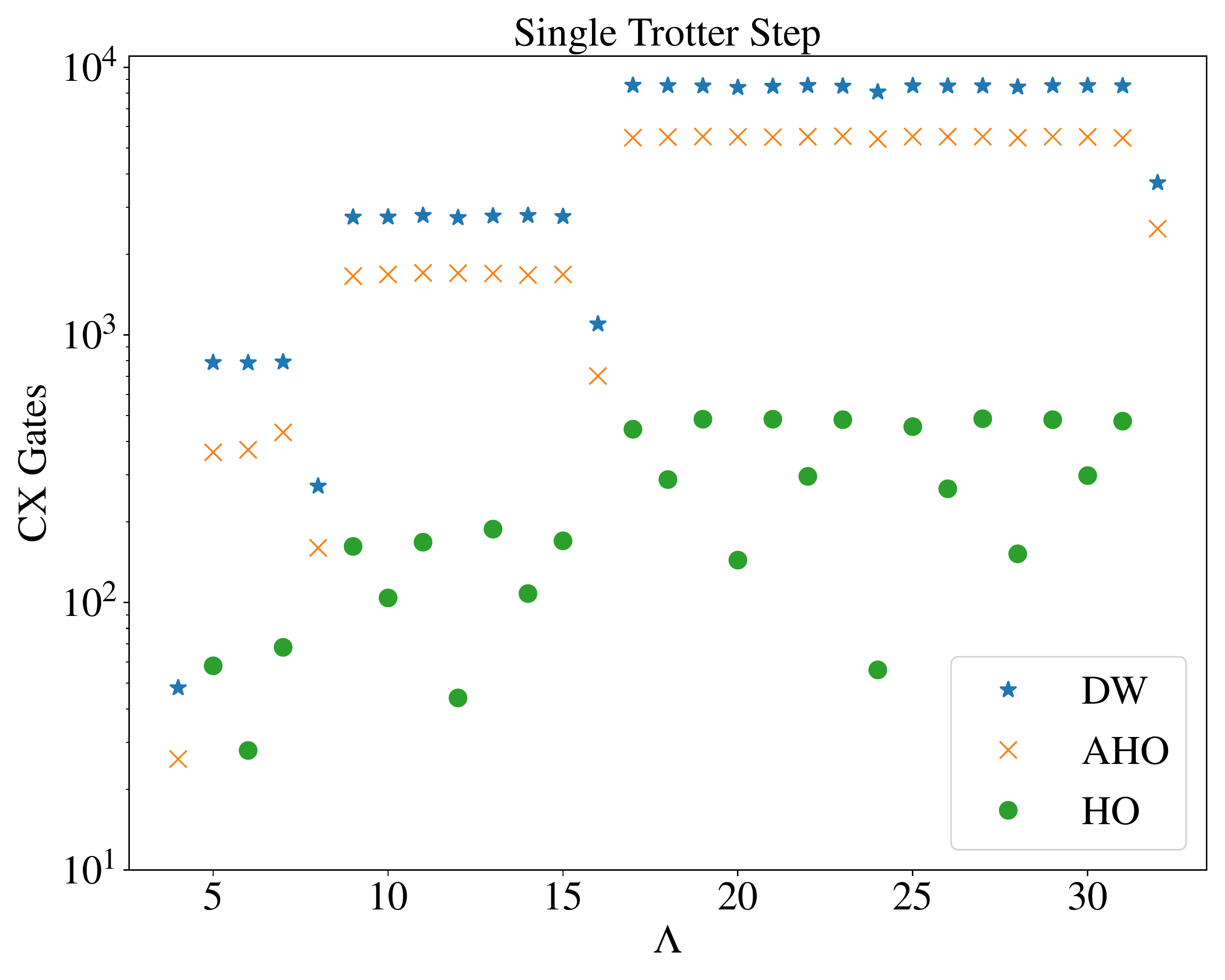}
    \caption{Number of entangling CX gates in a single Trotter step, as a function of the cutoff for the three superpotentials (with $m=g=\mu=1$) considered so far: the harmonic oscillator (green circles), the anharmonic oscillator (yellow crosses), and the double well (blue stars).  When the cutoff is equal to a power of two ($\Lambda = 2^n$) there are significant reductions in the number of CX gates required.  In particular, this produces zero CX gates for the harmonic oscillator, which isn't visible on this semi-log plot.}
    \label{fig:gate_counts}
\end{figure}

\mysubsection{Harmonic Oscillator}
The supersymmetric harmonic oscillator~(HO) is specified by the superpotential
\begin{equation}
    W(\qhat)=\frac{1}{2}m\qhat^2,
\end{equation}
where $m$ is the mass of the boson and fermion.  This is one of the simplest superpotentials to consider since $W''(\qhat)=m$ and the bosons and fermions do not interact.  No dynamical supersymmetry breaking is expected for this superpotential~\cite{Joseph:2020gdh}.
The energy spectrum as a function of the cutoff determined via exact diagonalization is plotted in Fig.~\ref{fig:ho_spectrum}.  To preserve supersymmetry the ground state-energy must vanish, and for any cutoff the value is exactly zero.  In fact, for this superpotential the only cutoff effect seems to be the lack of excited states; the energies of all accessible states are exact even for small cutoffs.
The results of the VQE for representative cutoffs up to $\Lambda \leq 32$ are given in Table~\ref{table:vqe}.  For this simple superpotential the VQE clearly converges to zero energy, confirming that a sufficiently low-noise quantum simulation would come to the correct conclusion about supersymmetry breaking.  For simulating the real-time evolution of this system, the entangling (CX) gate counts are given by the green circles in Fig.~\ref{fig:gate_counts}.  Notably, if the number of oscillator modes is set to a power of two, $\Lambda = 2^n$, no CX gates are required to evolve the system in time.  This makes the HO superpotential a promising benchmark for even today's quantum devices.

\begin{table}
\caption{\label{table:vqe}Minimum energy from 100 runs of the VQE for various supersymmetric superpotentials (with $m=g=\mu=1$).  The harmonic and anharmonic oscillator are expected to preserve supersymmetry and have a ground state energy of zero.  The double well breaks supersymmetry and should have a non-zero ground state energy.  The exact value represents the minimum energy from classical diagonalization.}
\begin{subtable}[t]{0.33\linewidth}
    \centering
    \begin{tabular}{c c c}
        \hline
        $\Lambda$ & Exact & VQE  \\
        \hline\hline
2 & 0.00e+00 & 5.34e-10 \\
4 & 0.00e+00 & 1.07e-09 \\
8 & 0.00e+00 & 4.06e-09 \\
16 & 0.00e+00 & 1.13e-08 \\
32 & 0.00e+00 & 4.81e-08 \\
          \hline
    \end{tabular}
    \caption{Harmonic oscillator.}

\end{subtable}%
\begin{subtable}[t]{0.33\linewidth}
    \centering
    \begin{tabular}{c c c}
        \hline
        $\Lambda$ & Exact & VQE  \\
        \hline\hline
2 & 9.38e-01 & 9.38e-01 \\
4 & 1.27e-01 & 1.27e-01 \\
8 & 2.93e-02 & 2.93e-02 \\
16 & 1.83e-03 & 6.02e-02 \\
32 & 1.83e-05 & 6.63e-01 \\
          \hline
    \end{tabular}
    \caption{Anharmonic oscillator.}
\end{subtable}%
\begin{subtable}[t]{0.33\linewidth}
    \centering
    \begin{tabular}{c c c}
        \hline
        $\Lambda$ & Exact & VQE  \\
        \hline\hline
2 & 1.08e+00 & 1.08e+00 \\
4 & 9.15e-01 & 9.15e-01 \\
8 & 8.93e-01 & 8.93e-01 \\
16 & 8.92e-01 & 8.94e-01 \\
32 & 8.92e-01 & 8.95e-01 \\
          \hline
    \end{tabular}
    \caption{Double well.}
\end{subtable}
\end{table}

\begin{figure}
    \centering
    \includegraphics[width=0.45\linewidth]{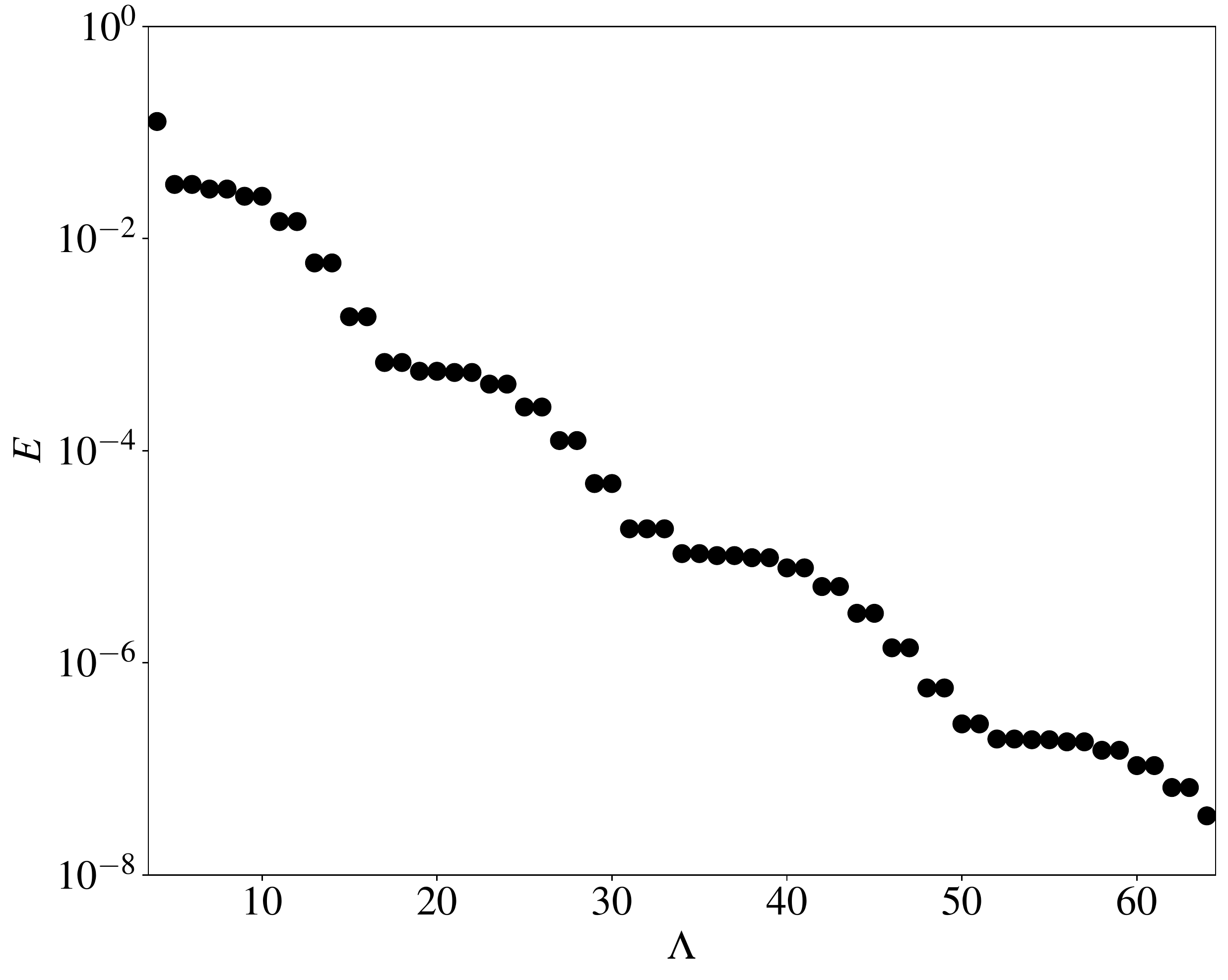}
    \caption{Semi-log plot of the ground-state energy of the anharmonic oscillator (with $m=g=1$) as a function of the cutoff.  As the cutoff increases this energy exponentially approaches zero signifying that supersymmetry is preserved in this model.}
    \label{fig:aho_spectrum_zoom}
\end{figure}

\mysubsection{Anharmonic Oscillator}
The supersymmetric anharmonic oscillator~(AHO) is given by the superpotential
\begin{equation}
    W(\qhat)=\frac{1}{2}m\qhat^2+\frac{1}{4}g\qhat^4,
\end{equation}
where $m$ is the mass of the boson and the coefficient $g$ dictates the strength of the interaction between the boson and fermion.  For any value of $g$ the AHO is also expected to preserve supersymmetry~\cite{Joseph:2020gdh}.  In Fig.~\ref{fig:ho_spectrum} we plot the low-lying energy spectrum as a function of the cutoff using exact diagonalization.  Here we see more significant errors induced by the cutoff, but as $\Lambda$ increases the low-lying energies rapidly converge.  In Fig.~\ref{fig:aho_spectrum_zoom} we plot the ground-state energy on a semi-log scale to show that it is approaching zero exponentially quickly as $\Lambda$ increases, which confirms that supersymmetry is preserved for this superpotential.  The results for the VQE in Table~\ref{table:vqe} show that it becomes difficult to find the expected ground-state energy for larger cutoffs~($\Lambda>16$).  This could be due to the fact that our wave function is not expressive enough, and we are currently exploring a greater variety of parameterizations.
The entangling gate count for a Trotter step is shown by the yellow crosses in Fig.~\ref{fig:gate_counts}.  Here we can see a dramatic increase in CX gates as a function of the cutoff compared with the HO superpotential.  There is still a significant decrease in the number of entangling gates when $\Lambda = 2^n$, although it is always non-zero.  This system will be very interesting to analyze using future NISQ hardware with improved error rates.

\begin{figure}
    \centering
    \includegraphics[width=0.45\linewidth]{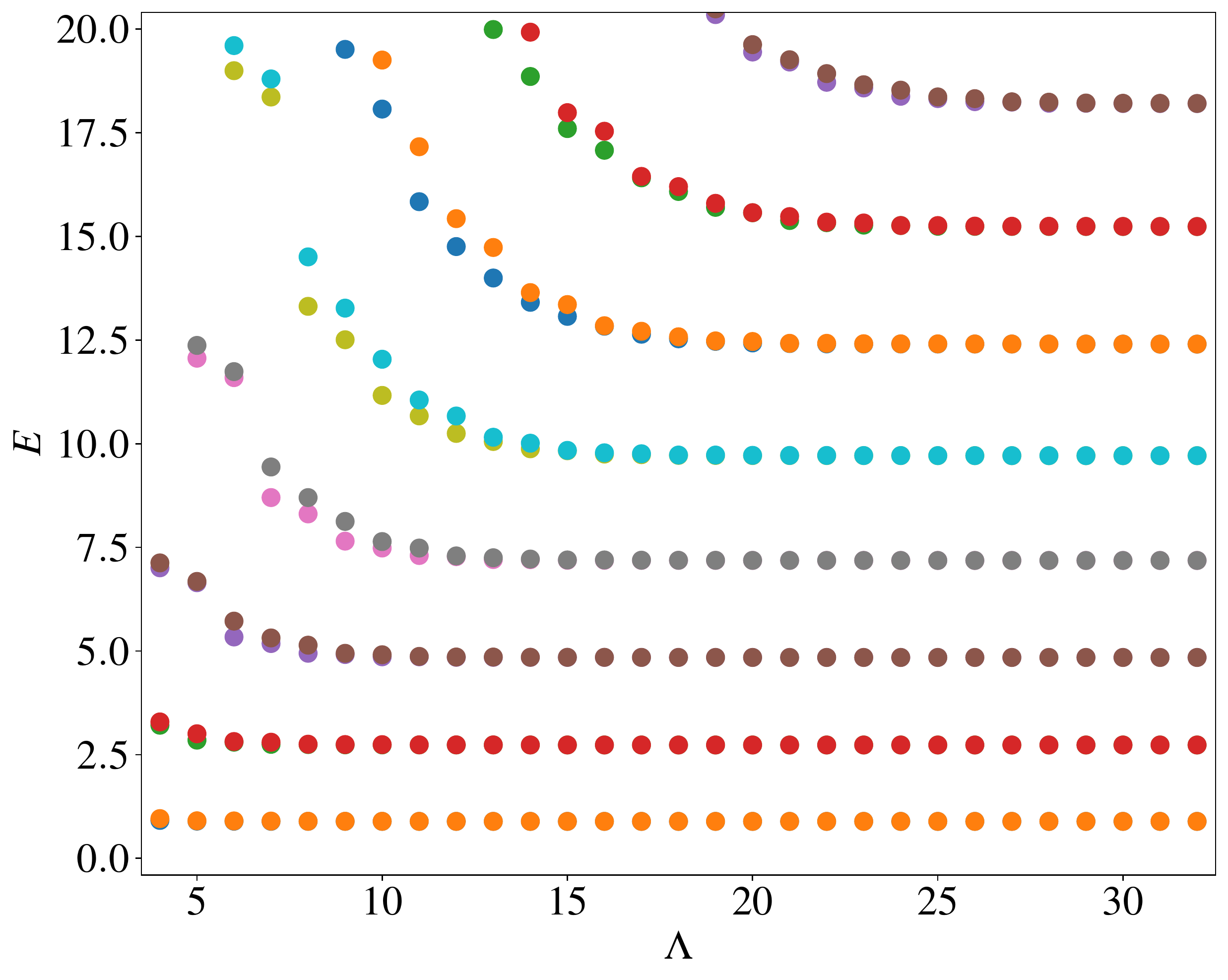}
    \caption{Similar to Fig.~\ref{fig:ho_spectrum} but for the double-well superpotential (with $m=g=\mu=1$).  As the large-cutoff limit is taken two-fold degeneracy is again recovered---now also for the ground-state energy, which is non-zero due to spontaneous supersymmetry breaking.}
    \label{fig:dw_spectrum}
\end{figure}

\mysubsection{Double Well}
The supersymmetric double well~(DW) is defined by the superpotential
\begin{equation}
    W(\qhat)=\frac{1}{2}m\qhat^2+g(\frac{1}{3}\qhat^3+\qhat\mu^2),
\end{equation}
where $m$ is the mass while the coefficients $g$ and $\mu$ control the strength of interactions between the boson and fermion.  For non-zero values of $g$ and $\mu$, this system is expected to exhibit dynamical supersymmetry breaking~\cite{Joseph:2020gdh}.  In Fig.~\ref{fig:dw_spectrum} we plot the spectrum from exact diagonalization as a function of the cutoff.  Similarly to the AHO, cutoff effects are non-negligible for very small values of $\Lambda$.  As the large-cutoff limit is taken the ground-state energy clearly does not converge to zero, confirming that supersymmetry is dynamically broken.  The VQE results in Table~\ref{table:vqe} successfully reproduce this non-zero ground-state energy.
The entangling gate count in Fig.~\ref{fig:gate_counts} scales in a comparable way to the AHO superpotential, with significant cost reductions for cutoffs equal to a power of two.

\mysection{Summary}
We have presented first results from our work using quantum computing to study supersymmetric quantum mechanics in $0+1$ dimensions.  As a measure of supersymmetry breaking we compute the ground-state energy using both classical exact diagonalization and the hybrid quantum--classical VQE.  Both methods confirm that supersymmetry is preserved for the harmonic oscillator and spontaneously broken for the double-well superpotential.  For the anharmonic oscillator exact diagonalization confirms the expected preservation of supersymmetry, while the VQE has difficulties that we are currently investigating by exploring whether different trial wave functions or classical optimizers can improve its performance.  Lastly we discussed the entangling gate counts for a single Trotter step for each of the three superpotentials, to illustrate why we consider supersymmetric quantum mechanics a promising target for quantum computing in the NISQ era.

\vspace{12 pt}
\noindent \textsc{Acknowledgments:}
We thank Yannick Meurice and Hank Lamm for encouraging discussions.
This work was supported by UK Research and Innovation Future Leader Fellowship {MR/S015418/1} and STFC grant {ST/T000988/1}.

\bibliographystyle{JHEP}
\bibliography{lattice21}
\end{document}